\documentclass[a4paper,german]{article}
\usepackage[affil-it]{authblk}
\usepackage[utf8]{inputenc}
\usepackage{graphicx,float}
\usepackage[backend=bibtex,sorting=none]{biblatex}
\bibliography{literatur} 

\begin{document}

\title{Universality and beyond in optical microcavity billiards with source-induced dynamics}

\author{Lukas Seemann  $^{1}$ and Martina Hentschel $^{2}$}

\affil{$^{1}$ \quad Institute of Physics, Technische Universit\"at Chemnitz, D-09107 Chemnitz, Germany; lukas.seemann@physik.tu-chemnitz.de\\
	$^{2}$ \quad Institute of Physics, Technische Universit\"at Chemnitz, D-09107 Chemnitz, Germany; martina.hentschel@physik.tu-chemnitz.de}



\maketitle

\begin{abstract}
Optical microcavity billiards are a paradigm of a mesoscopic model system for quantum chaos. We demonstrate the action and origin of ray-wave correspondence in real and phase space using far field emission characteristics and Husimi functions. Whereas universality induced by the invariant-measure dominated far field emission is known to be a feature shaping the properties of many lasing optical microcavities, the situation changes in the presence of sources that we discuss here. We investigate the source-induced dynamics and the resulting limits of universality while we find ray-picture results to remain a useful tool in order to understand the wave behaviour of optical microcavities with sources. We demonstrate the source-induced dynamics in phase space from the source ignition until a stationary regime is reached comparing results from ray, ray-with-phase, and wave simulations and explore ray-wave correpondence. 
\end{abstract}

\section{Introduction}

Two-dimensional (2D) system have inspired the field of quamtum chaos for many years \cite{Stoeckibuch, Haakebuch, Takabuch}. The origins of studying the quantum mechanical pendants of classically non-integrable systems trace back to the 1980ies when universality was established as a common property of very different chaotic systems \cite{Casati1980, Bohigas1984}, in particular in the energy level statistics in the 1980 paper by Casati et al. \cite{Casati1980}, and initiated a variety of studies focussing on the statistical properties based e.g. on Random Matrix Theory \cite{MehtaRMTbuch}.

The superb properties of this new class of mesoscopic model systems \cite{Imrybuch} for both electrons and photons soon initiated an interest in possible applications. Besides the ballistic quantum dots \cite{Takabuch}, the optical microcavities \cite{Vahala_book, QuaChaCav:Nockel:1997, QuaChaCav:Gmachl1998} received a lot of interest, lately also in mesoscopic and Dirac Fermion optics \cite{diracfermionoptics}. One pratical motivation was certainly the realization of microcavity lasers with directional emission, and plenty of solutions were found and investigated \cite{RMP_Wiersig_Cao}. One realization involves deformed microdisk cavities of various shapes \cite{shortegg}, including the Lima\c{c}on cavity \cite{limacon, limacon_Cao2009,limacon_Kim2009,limacon_Susumu_Taka2009,limacon_Yan2009,limacon_NJPhys,limacon_Albert2012}. Besides the experimental verification of the predicted \cite{limacon} directional and universal, resonance-independent far field emission originating from the cavity's invariant manifold, a remarkable ray-wave correspondence was seen. While all results were obtained for very differenct wavelengths $\lambda$ -- the ray modelling in the $\lambda\to 0$-limit, the wave simulation for $\lambda$ larger than the experimentally relevant values -- the agreement between all three aproaches was convincing with slight, interference inspired deviations between the three curves. Shinohara et al. \cite{limacon_Susumu_Taka2009} complemented this interpretation nicely by showing that averaging over a large number of resonances (42 in Ref.~\cite{limacon_Susumu_Taka2009}) improves ray-wave correpondence by averaging out the resonaance-specific features.

The reason for the universality of the observed far field emission properties is that the so-called natural measure (or Fresnel-weighted unstable manifold or steady probability
distribution) \cite{unstabmanif} determines the emission characterisitcs. Assuming light to be initially captured by total internal reflection, it will, in a chaotic cavity, violate this condition 
and its angle of incidence $\chi$ will cross the critical lines $\sin \chi_c = \pm 1/n$ in phase space ($n$ is refractive index of the cavity and we assume $n_0$=1 outside). This crossing of the critical line will be ruled by the unstable manifold of the system, weighted by the Fresnel reflection coefficient for our open, optical system -- as this quantity describes the expanding directions along which the light will escape the cavity (actually by evanescent escape in the wave picture). This implies that the unstable manifold, as an important and central, yet abstract quantity of nonlinear dynamics is directly accessible and visible in experiments and the corresponding simulations. We point out that therefore simulations of the passive, non-lasing cavity can succesfully describe even lasing cavities as long as mode interactions \cite{TakainCasati} do not play a role. In terms of ray picture modelling, the initial conditions are homogeneously distributed in phase space and the far field characteristics is recorded when an initial, transient regime is lapsed. 

\begin{figure}
	\centering
	\includegraphics[width=11cm]{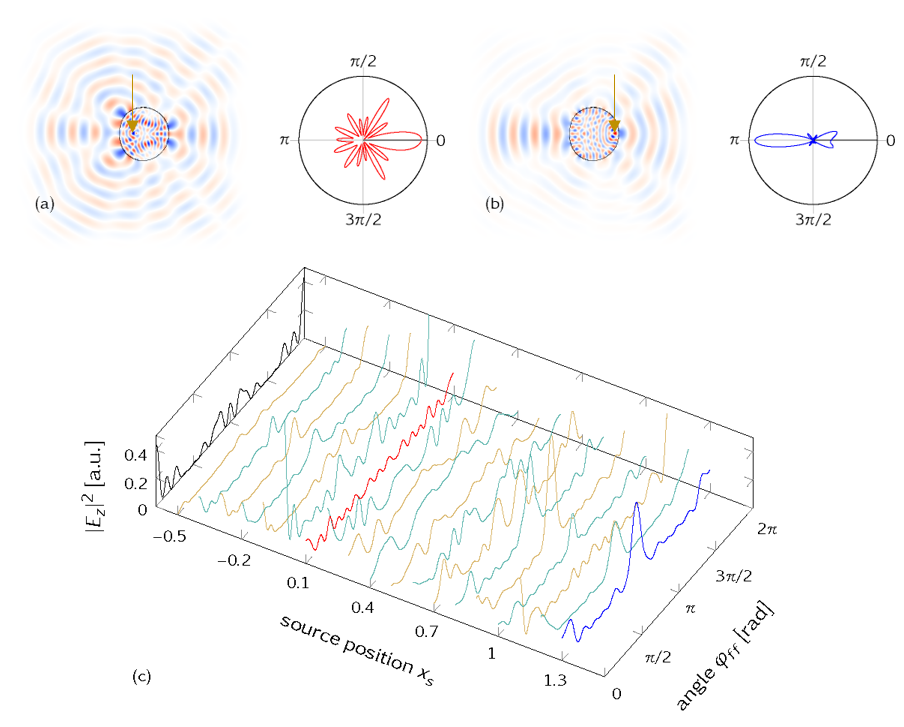}
	\caption{(a) Amplitude distribution $E_z$ in real space for a mode at resonant frequency $f=1.2$ and source position $x_s$=0.1 (marked by arrow), and corresponding far field emission as polar plot $\varphi_{ff}$. (b) Same for $x_s=1.3$. (c) Far field emission $|E_z|^2$ for source positions $x_s$ varied along the $x$-axis revealing a high sensitivity on $x_s$. The black line on the left marks the far field emission without source (uniform initial conditions).
	}
	\label{fig:farf_sourcepos_var}
\end{figure}

In this paper, we will explore another situation 
in optical microcavities that is induced by the presence of sources (or, similarly, relevant in microlasers with non-uniform pumping conditions \cite{HentschelKwon_spirallaser}). This setting can capture situations where not all initial conditions are homogeneously populated, in contrast to the microlaser case discussed above. Source-induced phenomena can be relevant, for example, due to a specific distribution of fluorescent particles, a local pumping scheme, or even due to the coupling of two or more systems that effectively changes the initial conditions to be non-uniform in phase space. The base element of any source can be described as a point-like emitter.

The paper is organized as follows. We will consider point-like sources and study their    
impact on the far field emission in chapter \ref{chap_sources}. We then investigate the source-initiated dynamics in phase space and introduce a ray picture extended by the phase information in chapter \ref{chap_dyn} before we end with a conclusion and summary in chapter \ref{chap_concl}.

\section{Optical microcavity billiards with sources}
\label{chap_sources}

\begin{figure}
	\centering
	\includegraphics[width=11cm]{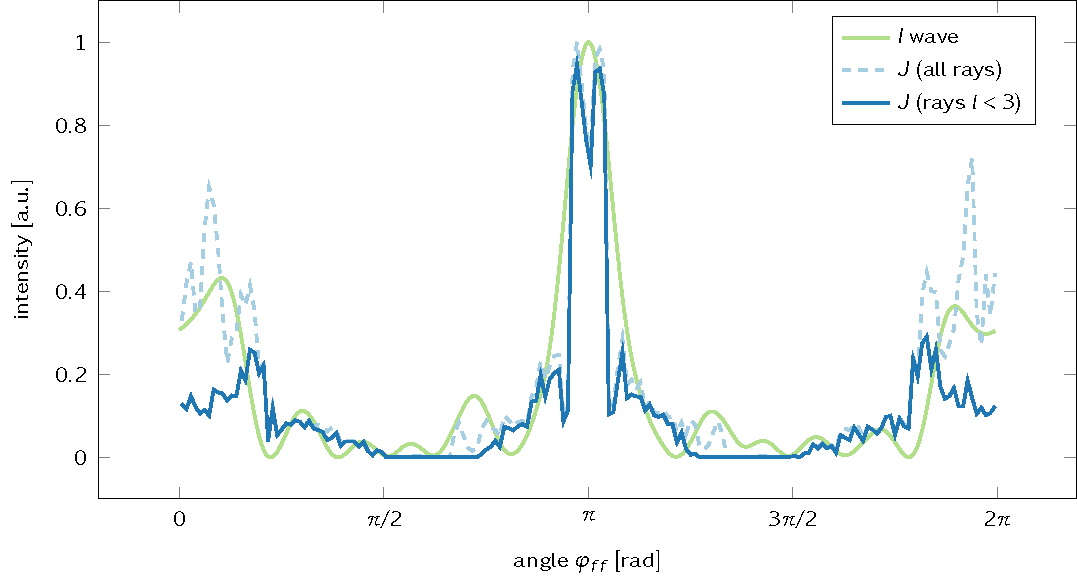}
	\caption{Far field emission for source position $x_s=1.3$. Wave intensity $I=|E_z|^2$ for the resonance at frequency $f=1.2$ (full green line) and for ray simulated intensities $J$ including the far field contribution of short rays only (full blue line) and of all rays (dashed line). Evidently, short rays with a trajectory length $l<3$ contribute significantly to the far field emission.}
	\label{fig:farf_raywave}
\end{figure}

We start our investigation for a Lima\c{c}on cavity \cite{limacon} with the shape given in polar coordinates $(r, \phi)$ as $r(\phi) = R_0 (1 + \epsilon \cos \phi)$ where we set $R_0=1$ and choose the deformation parameter $\epsilon=0.43$, such that the phase space of the cavity is known to be almost fully chaotic. We use Birkhoff coordinates, i.e.~the arclength along the boundary starting at its intersection with the positive $x$ axis, and the sine of the angle of incidence $\chi$ of light travelling inside the cavity to specify the position in phase space. The far field angle $\varphi_{ff}$ is measured mathematically positive with respect to the positive $x$ axis. We will consider TE polarized light (electric field transverse to the resonator plane, i.e. along the $z$ axis, $\vec{E} = E_z \vec{e_z}$), a refractive index $n=3.3$, and vary the position $x_s$ of the source along the $x$ axis. 
We use simulations with the open source software package meep \cite{meep} and so-called meep units with the velocity of light set to 1, such that frequency $f$ and (vacuum) wavelength $\lambda$ are reciprocal to each other, as is the period $T= 1/f$.

\textit{Variation of the source position.}
For a mode at resonance frequency $f=1.2$, the far field emission depends critically on the source position $x_s$ as is visible in Fig.~\ref{fig:farf_sourcepos_var}. In general, more central source positions relate to more isotropic emissions and the far field emission characteristics of the uniformly pumped cavity can be completely lost. 
Similar results were found in a study of graphene and optical billiards 
in Ref.~\cite{diracfermionoptics} where the importance of lensing effects in particular for single-layer graphene cavities was discussed.

Despite this deviation from the universality seen in the uniform pumping case, ray-wave correspondence still holds as illustrated in Fig.~\ref{fig:farf_raywave} for $x_s$=1.3. The far field wave intensity $I=|E_z|^2$ and the ray-simulated intensities agree reasonably well. In particular, we find the wave intensity $I$ to be reproduced by the short rays with trajectory lengths $l<3$ correposponding to typically very few reflection at the system boundary. In other words, in the presence of sources the far field is mainly determined by refractive escape of rays leaving the source and dwelling very few reflections in the cavity. This indicates the relevance of lensing effects when the cavity acts similar to a thick lens. However, longer rays are needed to establish a semiquantitative agreement with the wave result for all far field angles $\varphi_{ff}$. These long trajectory carry the information of the cavity geometry as a whole, namely in terms of the unstable manifold.

\begin{figure}
	\centering
	\includegraphics[width=11cm]{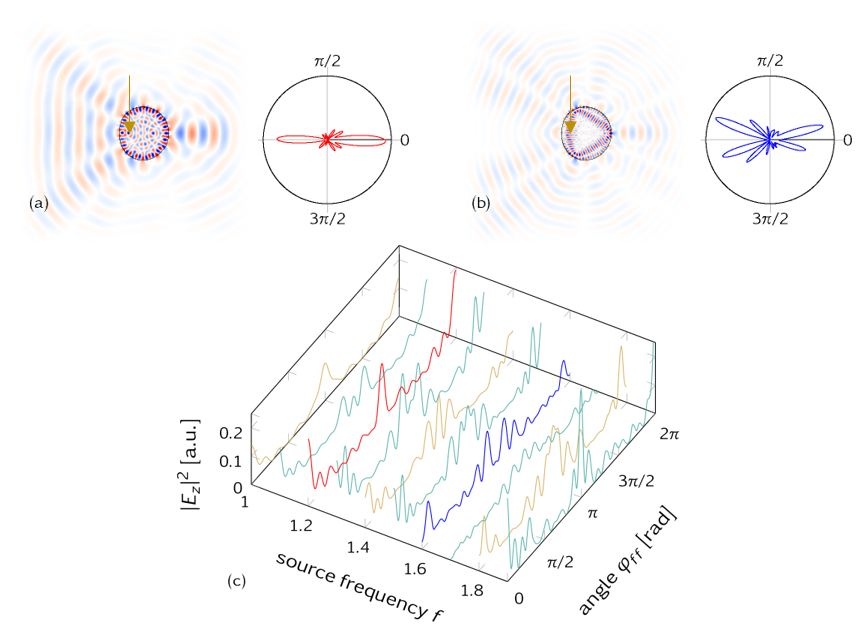}
	\caption{
		Far field emission depending on the source frequency f, while the source position is fixed at $x_s=-0.42$ (marked by arrow). 
		(a) At the resonance frequency $f$=1.2, the real space amplitude $E_z$ and the far field emission (polar plot) are shown. 
		(b) An off-resonant frequency $f=1.6$ yields a different mode pattern and a far field that differs in the details, but preserves the generic emission characteristics towards $\varphi_{ff} = 0$ and $\pi$. 
		(c) Comparison of different source frequencies $f$ confirms that the details are frequency dependent, while the overall far field emission is rather robust and dominated by the position of the source. 
	}
	\label{fig:farf_freq_var}
\end{figure}

\textit{Variation of the source frequency.}
It is worthwhile to characterize the far field sensitivity against variations of the source frequency, cf.~Fig.~\ref{fig:farf_freq_var}. In Fig.~\ref{fig:farf_freq_var}(a,b), mode patterns (amplitude $E_z$) are shown for two different frequencies $f$: resonant ($f=1.2$) in Fig.~\ref{fig:farf_freq_var}(a) and off-resonant ($f=1.6$) in  Fig.~\ref{fig:farf_freq_var}(b). The source position is fixed again on the $x$ axis, here at $x_s$=-0.42 (marked by arrows). While the intra-cavity patterns deviate a lot -- as expected upon a change of wavelength -- the emission characteristics is less affected, cf.~Fig.~\ref{fig:farf_freq_var}(c). 

This result may seem surprising at first sight. However, it can be straightforwardly interpreted on the basis of ray-wave correspondence. Our starting point is ray-wave correspondence for resonant frequencies as illustrated in Fig.~\ref{fig:farf_raywave} and confirmed in numberless other situations. As the naive ray picture does not know about frequencies or wavelengths, 
it would suggest no frequency dependence of the far field emission at all. Of course, this cannot be correct as we know that the details of the far field emission will be resonance, or more generally, frequency-dependent \cite{limacon_Susumu_Taka2009}. This is precisely what can be seen in Fig.~\ref{fig:farf_freq_var}(c).

\section{Source-induced dynamics: Ray-wave correspondence in phase space}
\label{chap_dyn}

So far we discussed the implications of the presence of sources inside billiards for light mainly in real space and in terms of far fields, and for the stationary situation. We will now complement the discussion in phase space, discussing both the Husimi function \cite{husimiEPL} and the ray signature of the source-induced dynamics. To this end we will discuss the dynamics initiated when a source is turned on and follow it until a stationary state (with the source constantly emitting) is reached. 

\begin{figure}
	\centering
	\includegraphics[width=11cm]{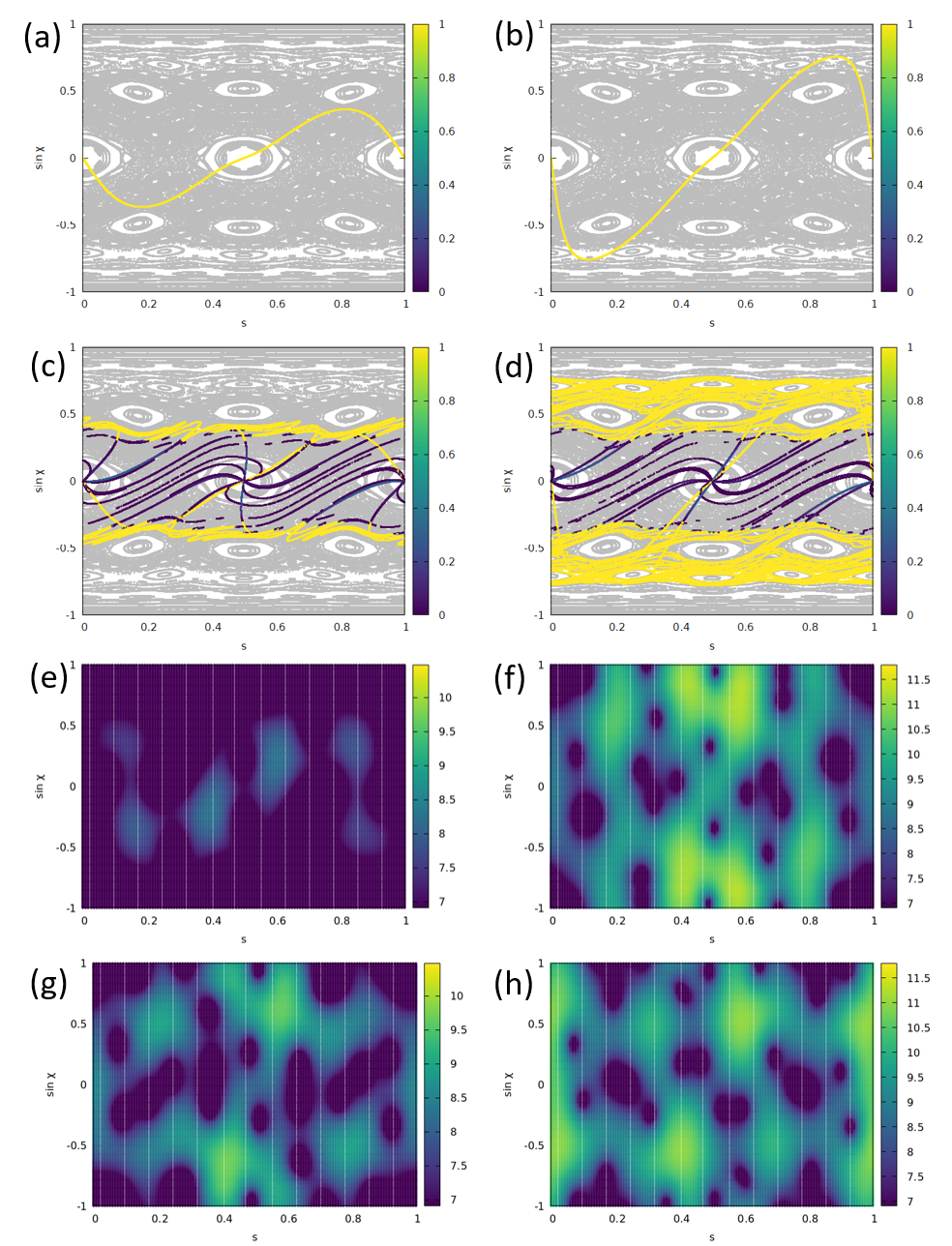}
	\caption{Dynamics induced by the source represented in phase space for two different source positions $x_s=0.6$ in (a,c,e,g) and $x_s=1$ in (b,d,f,h). In the ray simulation results (a-d), the Poincaré surface of section for a Lima\c{c}on cavity with $\epsilon=0.25$ is indicated as gray background. (a,b) Characteristicis of a homogeneously emitting source at the first boundary reflection. Shown is the Fresnel-weighted intensity inside the cavity. (c,d) Same as (a,b) but after 30 reflections when stationarity is reached. In (e-h), the wave-simulation results are visualized in phase space in terms of the Husimi function $H_{\mathrm{in}}^1$ when a stationary regime was reached. The Husimi patterns are evolving periodically with apprximately $T/2$, and typical patterns are shown at time frames (e) $t=16$, (f) $t=9$, (g) $t=37$, and (h) $t=30$. }
	\label{fig:dyn3}
\end{figure}

We will consider a Lima\c{c}on-shaped cavity with intermediate deformation parameter $\epsilon=0.25$ where a rich, mixed phase space is present, see the gray structure in Fig.~\ref{fig:dyn3}(a). We will compare two source positions on the $x$ axis, namely a rather central positions $x_s=0.6$ and an outer position $x_s=1.0$, thereby manipulating the excitability of WG-type modes and trajectories. We choose a source frequency of $f=0.64$ and use meep units as before, such that the period of the oscillation $T=1/f\approx 1.56$. We will consider 80 time steps per period $T$ and use the time frame number $t$ as our variable of time. The time to travel across the cavity, i.e.~to travel the optical distance $2 n R_0$, 
yields to be 6.6 in meep units, so it will take about 4.23 $T$ or approximately $t= 338$ frames (time steps) to travel the cavity's diameter.  

\begin{figure}
	\centering
	\includegraphics[width=11cm]{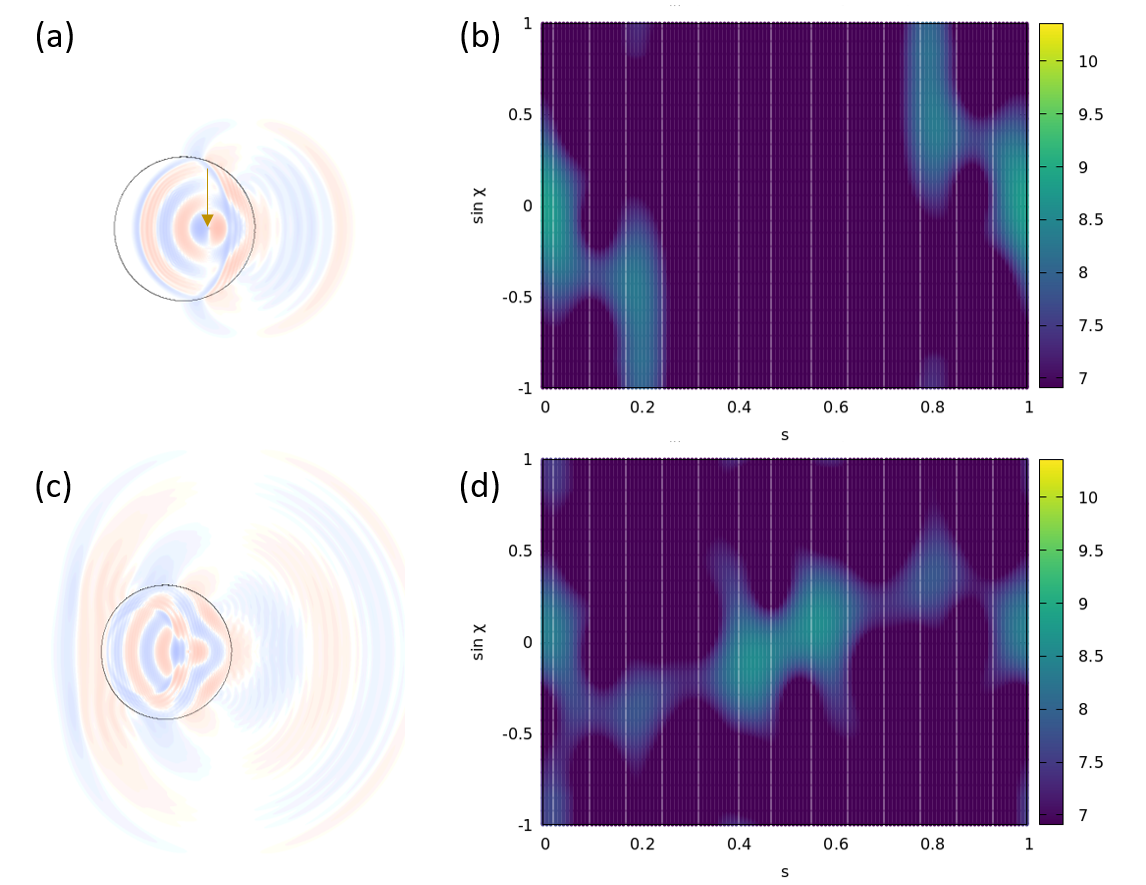}
	\caption{Early temporal evolution of an electromagnetic wave emitted from a source with $f=0.64$ at $x_s$=0.6 (marked by arrow in (a)), $\epsilon=0.25$.  (a) Real-space $E_z$ and (b) phase-space portrait in terms of the Husimi function $H_\mathrm{in}^1$ at time frame $t=186$, and similarly one period later at $t=272$ in (c) and (d). Note the extra signature in the center of (d) that can be attributed to the source signature after the first reflection at the boundary.  }   
	\label{fig:dyn1}
\end{figure}

{\textit{Initial dynamics.}}
What is the signature of the light emitted from the source? We start our study in the wave picture and excite a source placed at $x_s=0.6$ with a (resonant) frequency of $f=0.64$, cf.~Fig.~\ref{fig:dyn1}. The real space evolution of the electromagnetic field ampilitude $E_z$ is shown in Fig.~\ref{fig:dyn1}(a) for $t=186$, just before emitted light from the source reaches the far cavity interface. The corresponding phase-space representation is shown in Fig.~\ref{fig:dyn1}(b), and we use the incoming Husimi function $H_{\mathrm{in}}^1$ inside the cavity \cite{husimiEPL} to characterize its signature at the interface boundary where we will also take the Poincaré surface of section. We see that $H_{\mathrm{in}}^1$ contains the signature of light that has reached the cavity boundary at and around $s\approx 0$. 

The snapshots in Fig.~\ref{fig:dyn1}(c,d) are taken about one period $T$ later when all light emitted from the cavity at $t=0$ has reached the boundary. There is a distinct extra signature that must characterize light emitted from the source at its first reflection at the cavity interface where the Husimi function $H_{\mathrm{in}}^1$ is recorded (see also the yellow line in Fig.~\ref{fig:dyn3}(a) and the discussion there).

\begin{figure}
	\centering
	\includegraphics[width=11cm]{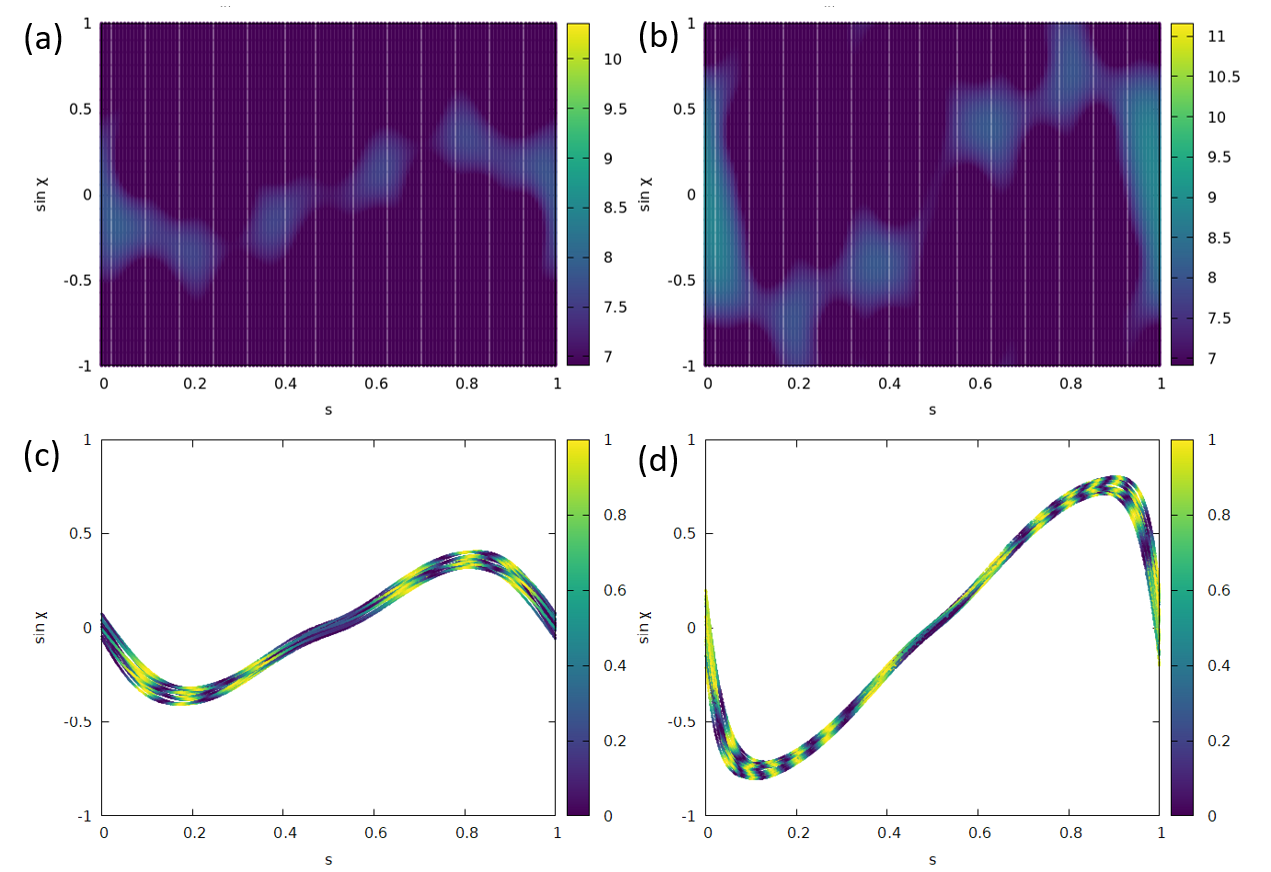}
	\caption{Phase-space representation of light emitted from a source in wave and ray-with-phase simulations. (a) Husimi function $H_{\mathrm{in}}^1$ for $x_s=0.6$ and $t= 253$. (b) $H_{\mathrm{in}}^1$ for $x_s=1.0$ and $t= 314$. (c) Ray-with-phase simulation of the intensity at the first reflection point, 20 sources were randomly placed in a square with side length 0.05 around $x_s=0.6$ and 2,500 rays were started isotropically from each source. (d) Same as (c) for $x_s=1.0$. }
	\label{fig:dyn2}
\end{figure}
This signature is discussed in more detail in Fig.~\ref{fig:dyn2} for two different source positions and in the wave and the phase-information extended ray picture, respectively. The ray's phase $\phi_p$ changing along the trajectory path is taken into account via $\phi_p = 2 \pi l/\lambda$ with $l$ the optical trajectory length travelled. Upon the reflection at the boundary, an additional phase shift would have to be taken into account, however, we will limit our study to just before the first reflection. 
Note that the wavelength enters the ray picture via the phase, and thus resonance-specific properties can, in principle, become accessible within the ray model. 

The Husimi functions shown in Fig.~\ref{fig:dyn2}(a,b) show comparable signatures and reach higher $|\sin \chi|$ for the outer source position $x_s=1.0$ in (c) as a direct geometrical consequence of $x_s$ being placed closer to the boundary. Notice that from frame to frame $t$, the location of the intensity maxima varies somewhat (not shown), indicating the importance of interference effects. 
These is straightforwardly confirmed qualitatively in Fig.~\ref{fig:dyn2}(c,d) where the phase-modulated ray intensity is shown at the first reflection point (in agreement with the time frames chosen for the Husimi plots) and clearly seen to possesss a rather similar structure.

{\textit{Stationary dynamics.}}
After having discussed the initial source dynamics, we will now consider the stationary state reached after the transient regime. The results are presented in Fig.~\ref{fig:dyn3}, again for the source positions $x_s=0.6$ (left column) and for $x_s=1.0$ (right column). 
We start our considerations in the naive ray picture (without phase information) for which we revisit the initial source dynamics in Fig.~\ref{fig:dyn3}(a,b). To this end 10,000 rays are started uniformly from the point-like source and traced to the first reflection with the boundary, yielding the yellow curves. We point out that the time needed to the first reflection point will depend on the starting direction of the ray, especially for non-central $x_s$. Although the reflection-number based Poincaré map representation will thus differ 
from the time frame $t$-based study used in the wave simulations, it still provides a useful tool that can be directly superimposed on the Poincaré surface of section (PSOS). The mixed structure of the PSOS is indicated as the gray background in Fig.~\ref{fig:dyn3}(a,b,c,d). 

In Fig.~\ref{fig:dyn3}(c,d) the source has been followed over several reflections at the boundary until stationarity (i.e.~intensity saturation) was reached. The phase-space distribution of the Fresnel-weighted intensity emitted by the source is indicated in color scale. The area between the critical lines $\sin\chi = \pm 1/n$ carries lower intensity due to refractive escape. It is evident that the spread in phase space depends on the source position - the closer to the boundary $x_s$ is, the larger $|\sin \chi|$ can be reached. In addition, we see that with the source positions chosen here, the 3-island orbits cannot be excited wihtin a ray simulation. 

The results of wave simulations in the stationary regime (after 50 $T$) are displayed in Fig.~\ref{fig:dyn3}(e-h). For both $x_s$, we find the pattern of the Husimi function $H_{\mathrm{in}}^1$ to periodically (with about $T/2$) vary. For each of the evolutions we pick two characteristic patterns. For $x_s=0.6$, we find a typical pattern that represents the source characteristics, cf.~Fig.~\ref{fig:dyn3}(e). Another one, cf.~Fig.~\ref{fig:dyn3}(g), displays intensity structured by the 3-island chains. Although these islands cannot be populated in the ray-based counterpart model, it may well be possible wihtin wave simulations due to a finite wavelength and when taking semiclassical corrections to the ray picture into account \cite{TureciStone02,Rex.2002,Hentschel_Schomerus_PRE2002,SchomerusHentschel_phase-space,Unterhinninghofen_PRE2008,Harayama.2015,PS_EPL2014,PS_JOpt2017}.

In particular, semiclassical arguments can explain why the intensity maxima in the Husimi function $H_{\mathrm{in}}^1$ seem to be placed at somewhat larger $|\sin \chi| $ in comparison to the ray model expectation. In the wave description, the evanescent wave associated with a WG-type mode will penetrate a distance of the order $\lambda$ into the outer space. The corrected ray picture analogue deploys the Goos-Hänchen shift that causes the reflection to take place at an effective interface \cite{GoosHaenchen,GoosHaenchen1949} such that the cavity appears effectively larger. But then the Husimi function is determined at a radius $R_0$ that is too small, thereby making the associated angle of incidence (somewhat) too large as illustrated in Ref.~\cite{Hentschel_Schomerus_PRE2002} and thus explaining the deviation. 

Eventually, for the other source position $x_s=1$, we illustrate two typical Husimi patterns in Fig.~\ref{fig:dyn3}(f,g). It is evident that now the $H_{\mathrm{in}}^1$ reaches larger values $|\sin \chi| $, in agreement with the ray-model expectation. One may speculate about the population of the 4-island orbit in Fig.~\ref{fig:dyn3}(f) or a mode beating interaction induced by the source. We will investigate this in further studies.

\section{Conclusion}
\label{chap_concl}

\begin{figure}
	\centering
	\includegraphics[width=11cm]{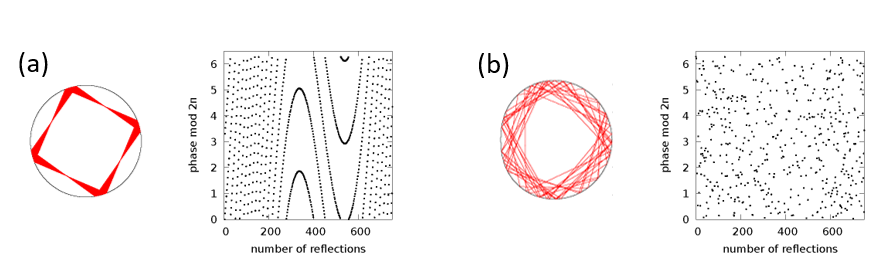}
	\caption{Phase evolution in a ray picture with phase for (a) a regular orbit ($\epsilon=0.25)$ and (b) a chaotic trajectory ($\epsilon=0.43$).}
	\label{fig:last}
\end{figure}

We have investigated optical microcavities in the presence of sources and discussed 
and explaind source-related (non-) universalities in the ray and wave pictures, and on the grounds of ray-wave correspondence. We find a nice agreement between the two approaches in terms of far fields and phase-space representations, and identified the signatures of the source in the ray and wave dynamics. We showed that a ray pciture extended by the phase information can improve the agreement by capturing interference effects and introducing a wavelength into the ray picture. The phase information can also be used to distinguish regular and chaotic orbits as is illustrated in Fig.~\ref{fig:last}: a chaotic trajectory can be associated with an unpredictable phase value at  the reflection points, whereas the phase evolves regularily for a periodic orbit.
However, a possible source-induced mode dynamics \cite{TakainCasati} that might be suggested by Fig.~\ref{fig:dyn3} in the stationary regime is beyond the scope of ray-wave correspondence.

\section{Acknowledgements}

We thank Tom Rodemund and Marika Carmen Federer for discussions. The special thanks of M.H. goes to Giulio Casati, whom she had the chance to meet several times as a young PhD student when the very basics of this work were laid. 


\printbibliography

\end{document}